\documentclass[prb,showpacs]{revtex4}
\usepackage{graphicx}

\newcommand{\ket}[1]{\left\vert#1\right\rangle}
\newcommand{\bra}[1]{\left\langle#1\right\vert}
\newcommand{\abs}[1]{\left\vert#1\right\vert}
\def\real{\mathop{\rm Re}\nolimits}     

\begin{document}

\title{A(nother) Continuum Model for Dephasing in Mesoscopic Systems}
\author{S. \c{S}enozan}
\author{S. Turgut}
\author{M. Tomak}
\affiliation{
Department of Physics, Middle East Technical University,\\
06531, ANKARA, TURKEY}
\date{\today}

\begin{abstract}
A dephasing model in the spirit of B\"uttiker's fictitious probe
model where infinite probes are distributed uniformly over the
conductor is proposed. The dephasing rate enters into the model as
an adjustable parameter and to compute the conductance. A
one-dimensional delta function scatterer model is solved
numerically. We observe the dephasing effects on the calculated
conductance.
\end{abstract}

\maketitle


\section{INTRODUCTION}

Dephasing, the loss of coherence in wavefunction, is an important
phenomenon in mesoscopic systems. It is the phenomenon that
distinguishes the microscopic where full quantum coherence is the
rule and the macroscopic where there is no trace left of the quantum
phase. In the intermediate mesoscopic regime, its effect is
important. It is either due to the collisions with the other
electrons and phonons, which can be adjusted by temperature or it
can be influenced by external factors \cite{datta}.

Several models have been proposed for modelling the dephasing
effects, coherent absorption, wave
attenuation\cite{BenjaminJayannavar}, introducing random phase
fluctuations in the scattering matrix\cite{RandomPhase} are a few.
One of the oldest models is the fictitious probe model of
B\"uttiker.\cite{ButTemel,ButIBM} This model has been applied into
several different problems. Also, it has been changed as a model to
overcome some of its deficiencies; for example momentum
randomization is eliminated and pure coherence effects are brought
to front\cite{PureDephasing} and long stub model is applied for
satisfying the charge conservation requirement for time dependent
currents.\cite{BeenakkerStub} (But the stub model is introduced
earlier). The model can be justified based on microscopic
theory.\cite{Micro1,Micro2,Micro3} They are also generalized to the
continuous case where infinite probes are distributed continuously
over the conductor\cite{Micro1,Micro2,Micro3,Cont1,Cont2,Cont3}.

In this contribution, we are going to propose another model based on
B\"uttiker's fictitious probe model where infinite probes are
distributed continuously over the conductor.The inelastic scatterers
are modelled in terms of a scattering matrix with a coupling
parameter D, which sets the strength of the decoherence introduced.
The aim of this paper is to present this continuous model in order
to get the conductance of a one-dimensional conductor. In the next
section we define the discrete model and after that section the
continuum version of it and numerical procedure is given.The last
section is devoted to our results and conclusions.

\section{The Discrete Model}

\begin{figure}
\begin{center}
\includegraphics[scale=0.70]{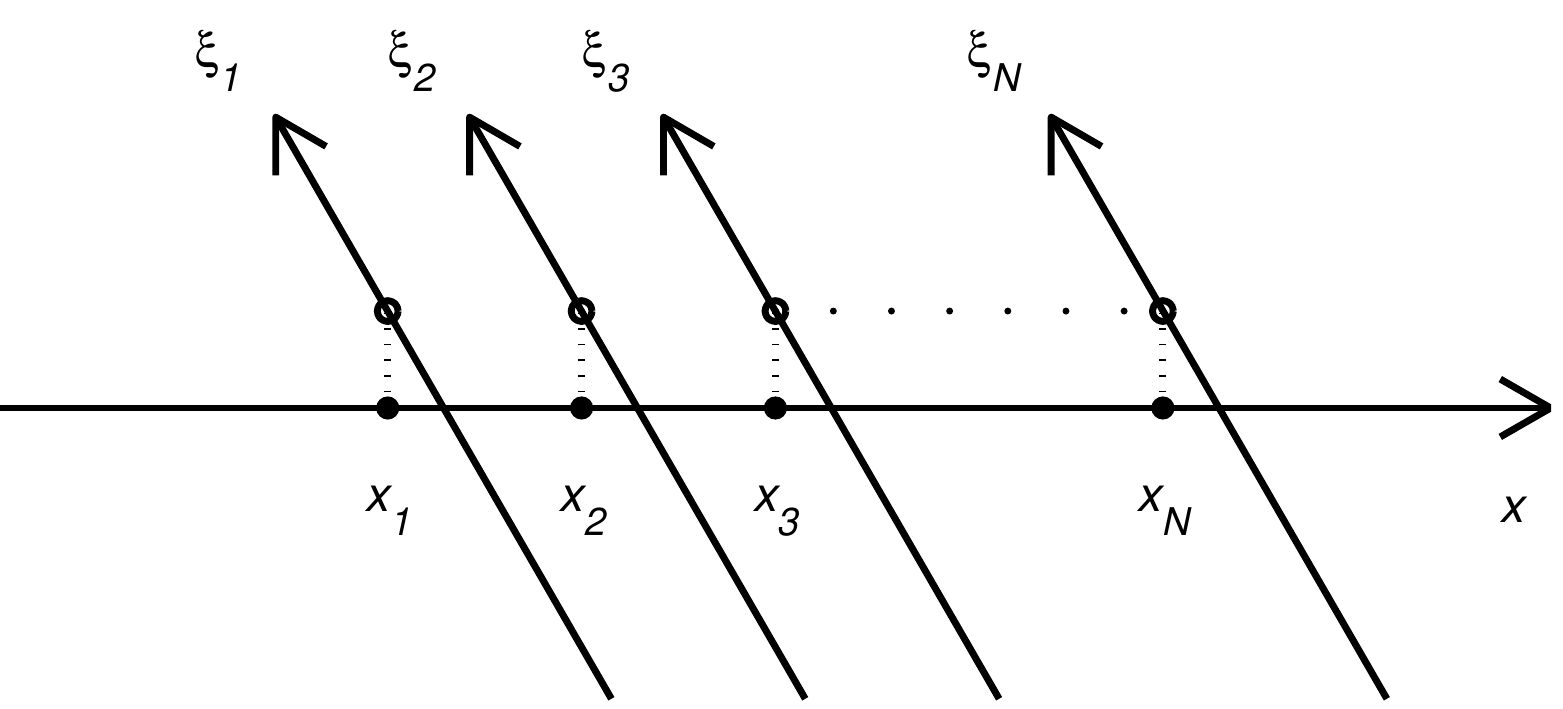}
\end{center}
\caption{The geometry of the problem.} \label{fig:geofprob}
\end{figure}

We are interested in extending B\"uttiker's model for decoherence in
1D transport in a way that decoherence proceeds at every location.
The geometry of the problem is shown in Fig.~\ref{fig:geofprob}.
Here there is a conductor along which electrons move and scatter.
Apart from that, $N$ additional probes are also placed for modelling
the decoherence effects on the main conductor. It is assumed that
the electrons can jump between the conductor and the probes. It can
go to equilibrium in those probes but will eventually return back
and at the end coherence with the wavefunction in the main conductor
will be lost.

In order to describe the possible states of the electrons, the state
of electron at position $x$ on the main conductor is denoted by
$\ket{x}$ and the state when the electron is on probe-$j$ at the
position $\xi$ will be denoted by $\ket{\xi,j}$. Any state
$\ket{\psi}$ can be expressed as a superposition of these as
\begin{equation}
\ket{\psi}=\int dx \psi(x)\ket{x}+\sum_j\int d\xi
\phi_j(\xi)\ket{\xi,j}\quad,
\label{eq_psi}
\end{equation}
where $\psi(x)$ is the wavefunction on the main conductor and
$\phi_j(x)$ is the wavefunction on probe-$j$. We let the potential
on the main conductor be $V(x)$. On the probes, we assume that the
electrons move freely, feeling the constant potential $V_j$ on
probe-$j$.

The Hamiltonian for the electrons is taken as
\begin{equation}
 H=h_0+\sum_j h_j +\sum_j d_j\left(\ket{\xi=0,j}\bra{x_j}+\ket{x_j}\bra{\xi=0,j}\right)\quad.
\end{equation}
where $h_0$ and $h_j$ denote those Hamiltonians of the main
conductor and probes  respectively and $d_j$ is a real number
representing the coupling strength to probe-$j$.

We can write down the following differential and abstract
representations of $h_0$ and $h_j$
\begin{eqnarray}
 h_0 &=-\frac{\hbar^2}{2m^*}\frac{d^2}{dx^2}+V(x)\quad,\quad
   h_0 &= \int\int dx dx^\prime h_0(x;x^\prime)\ket{x}\bra{x^\prime}\quad,\\
 h_j &=-\frac{\hbar^2}{2m^*}\frac{d^2}{d\xi^2}+V_j\quad,\quad
   h_j &= \int\int d\xi d\xi^\prime h_j(\xi;\xi^\prime)\ket{\xi,j}\bra{\xi^\prime,j}\quad.
\end{eqnarray}
Note that these operators act on their respective spaces. As a
result, we have $h_0\ket{\xi,j}=h_j\ket{x}=0$. Also
$h_j\ket{\xi,i}=0$ if $i\neq j$. As a result, for the state given in
Eq.~(\ref{eq_psi}) we have
\begin{eqnarray}
  h_0\ket{\psi} &=& \int dx \left(-\frac{\hbar^2}{2m^*}\frac{d^2\psi(x)}{dx^2}+V(x)\psi(x)\right)\ket{x}\quad,\\
  h_j\ket{\psi} &=& \int d\xi \left(-\frac{\hbar^2}{2m^*}\frac{d^2\phi_j(\xi)}{d\xi^2}+V_j\phi_j(\xi)\right)
    \ket{\xi,j}\quad.
\end{eqnarray}

In the Hamiltonian we add a term for the transfer of electrons
between probes and the main conductor. We assume that when the
electron is at position $x_j$ on the main conductor, it can jump to
the origin, $\xi=0$, of probe-$j$. The term in the Hamiltonian of
the form $\ket{\xi=0,j}\bra{x_j}$ handles this. The hermitian
conjugate handles the opposite process, namely jumping from
probe-$j$ to the main conductor.

Here $d_j$ could have been chosen complex valued, but this is
unnecessary since it does not introduce any new effects. Moreover,
the reality implies a simple time-reversal operation (complex
conjugation of wavefunction) and the symmetry implies that the
scattering matrix is symmetric.

The Schr\"odinger's equation, $H\ket{\psi}=E\ket{\psi}$ can be
expressed in terms of wavefunctions as
\begin{eqnarray}
\left(-\frac{\hbar^2}{2m^*}\frac{d^2\psi(x)}{dx^2}+V(x)\psi(x)\right)+\sum_j d_j \phi_j(0)\delta(x-x_j) &=&  E\psi(x)\quad,\\
\left(-\frac{\hbar^2}{2m^*}\frac{d^2\phi_j(\xi)}{d\xi^2}+V_j\phi_j(\xi)\right)
+ d_j\psi(x_j)\delta(\xi) &=& E\phi_j(\xi)\quad.
\end{eqnarray}
where we assume that the potential on the main conductor, $V(x)$, is
constant outside a certain interval.
\begin{equation}
 V(x) = \left\{
\begin{array}{ll}
V_L & \textrm{for}~~x<x_L^{(b)}\\
V_R & \textrm{for}~~x>x_R^{(b)}
\end{array}
\right.
\end{equation}
where between the points $x_L^{(b)}$ and $x_R^{(b)}$, $V(x)$ varies.
The scattering region and the points $x_j$ are contained in this
interval.

We will define the incoming wave amplitudes $a_j$, $a_j^\prime$ and
the outgoing wave amplitudes $b_j$, $b_j^\prime$ ($j=0,1,\ldots,N$)
for any solution at energy $E$ by
\begin{eqnarray}
 \psi(x) &=& \left\{
\begin{array}{ll}
\frac{1}{\sqrt{v_L}} \left( a_0 e^{ik_Lx} + b_0 e^{-ik_Lx}\right)  & \textrm{for}~~x<x_L^{(b)}\\
\frac{1}{\sqrt{v_R}} \left( a_0^\prime e^{-ik_Rx} + b_0^\prime
e^{ik_Rx}\right)  & \textrm{for}~~x>x_R^{(b)}
\end{array}
\right. \\
 \phi_j(\xi) &=& \frac{1}{\sqrt{v_j}}\left\{
\begin{array}{ll}
 a_j e^{ik_j\xi} + b_j e^{-ik_j\xi}  & \textrm{for}~~\xi<0\\
 a_j^\prime e^{-ik_j\xi} + b_j^\prime e^{ik_j\xi}  & \textrm{for}~~\xi>0
\end{array}
\right.
\end{eqnarray}
where for any energy $E$, the left and right wavenumbers are defined
as
\begin{equation}
k_L=\sqrt{\frac{2m^*(E-V_L)}{\hbar^2}}\quad,\qquad
k_R=\sqrt{\frac{2m^*(E-V_R)}{\hbar^2}}\quad. \end{equation}

 For the probe-$j$,
the electrons move freely with wavenumbers
\begin{equation}
 k_j=\sqrt{\frac{2m^*(E-V_j)}{\hbar^2}}\quad.\end{equation}
The corresponding velocities are defined accordingly, $v_L=\hbar
k_L/m^*$ etc.

There are $2N+2$ independent solutions of the wave equation. Any
particular solution can be obtained by choosing arbitrary values for
the incoming wave amplitudes $a_j$ and $a_j^\prime$. From these
values alone, the outgoing wave amplitudes $b_j$ and $b_j^\prime$
can be determined. The relation between the outgoing and incoming
amplitudes involves the scattering matrix
\begin{eqnarray}
  b_j &=& \sum_{i=0}^N S_{ji} a_i + S_{ji^\prime} a_i^\prime\quad,\\
  b_j^\prime &=& \sum_{i=0}^N S_{j^\prime i} a_i + S_{j^\prime i^\prime} a_i^\prime\quad.
\end{eqnarray}
Our purpose is to obtain the scattering matrix. Through this we can
calculate the transport properties of the system.

\subsection{Solution for probe-$j$}

First we write down the solution of the Schr\"odinger's equation for
probe-$j$. The wavefunction $\phi_j(\xi)$ is continuous at the
origin $\xi=0$, but its derivative has a discontinuity
\begin{equation}
 \Delta \phi_j^\prime(0) =\phi_j^\prime(0+)-\phi_j^\prime(0-)=\frac{2m^* d_j}{\hbar^2} \psi(x_j)\quad.\end{equation}
The outgoing amplitudes then can be expressed as
\begin{eqnarray}
  b_j^\prime &=& a_j - i D_j \psi(x_j)\quad,\label{eq_bj1}\\
  b_j &=& a_j^\prime - i D_j \psi(x_j)\quad,\label{eq_bj2}
\end{eqnarray}
where
\begin{equation}
   D_j = \frac{d_j}{\hbar\sqrt{v_j}}\quad.
\end{equation}
Since $d_j$ has dimensions Energy$\times$Length, $ D_j$ has the
dimensions of square root of velocity. We will need the following
expression below.
\begin{equation}
  \phi_j(0)= -\frac{i}{\sqrt{v_j}}\theta_j\quad,
\end{equation}
where
\begin{equation}
  \theta_j= D_j \psi(x_j) +  i(a_j+a_j^\prime)\quad.
\end{equation}

\subsection{Solution for the main conductor}

Schr\"odinger's equation for the main conductor can be expressed as
\begin{equation}
 [E-h_0]\psi(x) = \sum_j d_j \phi_j(0) \delta(x-x_j)=-i\hbar \sum_j  D_j \theta_j \delta(x-x_j)\quad.
\end{equation}
which can be solved easily by using the Green function as
\begin{equation}
\psi(x)=\psi_0(x) + \int dy G(x;y)\left(-i\hbar\sum_j  D_j \theta_j
\delta(y-x_j)\right)\quad, \label{eq_psimain}
\end{equation}
where $\psi_0$ is a particular solution of the homogeneous equation,
$[E-h_0]\psi_0=0$, and $G(x;y)$ is the Green function satisfying
\begin{equation}
 [E-h_0(x)]G(x;y)=\delta(x-y)\quad.\end{equation}

The wavefunction $\psi(x)$ is
\begin{equation}
 \psi(x) =a_0\varphi_L(x)+a_0^\prime \varphi_R(x) -i\hbar\sum_j G^{(+)}(x;x_j) D_j \theta_j \quad.\end{equation}
where $\varphi_L(x,E)$ and $\varphi_R(x,E)$ are two scattering
solutions of the main conductor. The general solution of the
homogeneous equation can be expressed as a superposition of these
two. These solutions satisfy
\begin{eqnarray}
 \varphi_L(x,E) &=& \left\{
\begin{array}{ll}
\frac{1}{\sqrt{v_L}} \left( e^{ik_Lx} + r_0 e^{-ik_Lx}\right)  & \textrm{for}~~x<x_L^{(b)}\\
\frac{1}{\sqrt{v_R}} ~t_0 e^{ik_Rx} & \textrm{for}~~x>x_R^{(b)}
\end{array}
\right. \\
 \varphi_R(x,E) &=& \left\{
\begin{array}{ll}
\frac{1}{\sqrt{v_L}}  ~t_0^\prime e^{-ik_Lx}  & \textrm{for}~~x<x_L^{(b)}\\
\frac{1}{\sqrt{v_R}} \left( e^{-ik_Rx} + r_0^\prime e^{ik_Rx}\right)
& \textrm{for}~~x>x_R^{(b)}
\end{array}
\right.
\end{eqnarray}
These are the solutions of $[h_0-E]\varphi_{L,R}=0$ obtained when
there are no probes connected. Here $r_0$, $r_0^\prime$, $t_0$ and
$t_0^\prime$ are reflection and transmission amplitudes and we have
$t_0=t_0^\prime$ due to the symmetry of the scattering matrix. Green
functions can be expressed in terms of these solutions,
$\varphi_{L,R}$. Note that in Eq.~(\ref{eq_psimain}), the term
containing the Green function can have only outgoing waves if
$G^{(+)}$ is used. In that case, all incoming waves should appear in
$\psi_0$. As a result we have $\psi_0=a_0\varphi_L+a_0^\prime
\varphi_R$.

Since $\theta_j$ depends on $\psi(x_j)$, we need to solve this
equation. To simplify the notation we first define $\theta_{0j}$ as
\begin{equation}
 \theta_{0j} =  D_j \psi_0(x_j)+i(a_j+a_j^\prime)\quad,\end{equation}
and note that $\theta_{0j}$ depends only on incoming wave
amplitudes. Using this, we get the following set of $N$ equations,
\begin{equation}
 \theta_\ell=\theta_{0\ell}-i\hbar\sum_j  D_\ell G^{(+)}(x_\ell,x_j)  D_j \theta_j \quad.\end{equation}
Let us now define an $N\times N$ matrix $\Gamma_{\ell j}$ as
\begin{eqnarray}
\Gamma_{\ell j} &=& \delta_{\ell j} +i\hbar  D_\ell
G^{(+)}(x_\ell,x_j)  D_j
        = \delta_{\ell j} +\frac{ D_\ell  D_j}{t_0} \varphi_R(x_{j_<}) \varphi_L(x_{j_>}) \\
        &=&  \delta_{\ell j}+\frac{1}{t_0} f_{Rj_<} f_{Lj_>}\quad,
\end{eqnarray}
where $f_{Rj}= D_j \varphi_R(x_j)$ and $f_{Lj}= D_j\varphi_L(x_j)$.
The final solution is
$\theta_j=\sum_\ell\left(\Gamma^{-1}\right)_{j\ell}\theta_{0\ell}$
from which we obtain all scattering amplitudes.

It may be shown that the inverse of $\Gamma$ can be written as
\begin{equation}
 \left(\Gamma^{-1}\right)_{\ell j}=\delta_{\ell j} - \frac{1}{t_d}
    ~\tau_{Rj_<}~\tau_{Lj_>}\quad,
\end{equation}
where
\begin{equation}
 \tau_L=\Gamma^{-1}f_L\quad,\quad \tau_R=\Gamma^{-1} f_R\quad,\quad
t_d=t_0-f_R^T\Gamma^{-1}f_L\quad.
\end{equation}

\subsection{The scattering matrix}
We look at the behavior of $\psi(x)$ for $x<x_L^{(b)}$.
\begin{eqnarray}
\psi(x)&=&\psi_0(x) -i\hbar\sum_{j\ell}  G^{(+)}(x;x_j) D_j\left(\Gamma^{-1}\right)_{j\ell}\theta_{0\ell}\\
    &=&\frac{1}{\sqrt{v_L}}a_0e^{ik_Lx}+\frac{1}{\sqrt{v_L}}\left(r_0a_0+t_0a_0^\prime - \sum_\ell \tau_{L\ell}\theta_{0\ell}\right)e^{-ik_Lx}
\end{eqnarray}
Therefore we have
\begin{equation}
 b_0 = r_0a_0+t_0a_0^\prime - \sum_\ell \tau_{L_\ell}\theta_{0\ell}\quad.\end{equation}
For $x>x_R^{(b)}$ we get
\begin{equation}
 b_0 = t_0a_0+r_0^\prime a_0^\prime - \sum_\ell \tau_{R_\ell}\theta_{0\ell}\quad.\end{equation}
The equations (\ref{eq_bj1},\ref{eq_bj2}) give the outgoing
amplitudes at the probes as follows
\begin{equation}
 b_j = - a_j -i \left(\Gamma^{-1}\right)_{j\ell}\theta_{0\ell}\quad,\qquad
b_j^\prime = - a_j^\prime -i \left(\Gamma^{-1}\right)_{j\ell}\theta_{0\ell}
\end{equation}
Finally, $\theta_{0\ell}$ depends only on the incoming wave amplitudes through
\begin{equation}
 \theta_{0\ell} = a_0 f_{L\ell}+a_0^\prime f_{R\ell} + i(a_\ell+a_\ell^\prime)\quad.\end{equation}

From these expressions we can read off the scattering matrix
elements as follows. First scattering amplitudes for the main
conductor
\begin{eqnarray}
  t_d=S_{LR}=S_{RL} &=& t_0 -f_L^T\Gamma^{-1}f_R=t_0-\tau_L^Tf_R=t_0-\tau_R^Tf_L\quad,\\
  S_{LL} &=& r_0 - f_L^T\Gamma^{-1}f_L=r_0-\tau_L^Tf_L\quad,\\
  S_{RR} &=& r_0^\prime - f_R^T\Gamma^{-1}f_R=r_0^\prime - \tau_R^Tf_R\quad.
\end{eqnarray}
We will use the symbol $t_d$ for the amplitude $S_{LR}$ and call
it the direct transmission amplitude. For the scattering into and
between the probes we have
\begin{eqnarray}
S_{Lj}=S_{Lj^\prime}=S_{jL}=S_{j^\prime L} &=& - i \tau_{Lj}\quad,\\
S_{Rj}=S_{Rj^\prime}=S_{jR}=S_{j^\prime R} &=& - i \tau_{Rj}\quad,\\
S_{j\ell}=S_{j^\prime\ell^\prime} &=& -\delta_{j\ell}+\left(\Gamma^{-1}\right)_{j\ell}\quad,\\
S_{j\ell^\prime}=S_{j^\prime\ell} &=& \left(\Gamma^{-1}\right)_{j\ell}\quad.
\end{eqnarray}
Note that $j$ and $j^\prime$ denote the negative and positive axes
respectively on probe-$j$. These two directions are entirely
equivalent for scattering. Therefore if an inversion is taken on
probe-$j$ (i.e., $j$ is switch with $j^\prime$) then the
scattering matrix should remain invariant. This symmetry can be
seen in the expressions above.

For example, for the scattering between two different probes $j$
and $\ell$ ($j\neq\ell$), the scattering amplitude is
\begin{equation}
 -\frac{1}{t_d} ~ \tau_{Rj_<} ~ \tau_{Lj_>} \end{equation}
independent of the directions the wave comes and goes. If a wave
coming from probe-$j$ is scattered back into the same probe
(perhaps through passing to the main conductor) then the
transmission amplitude is
\begin{equation}
 (\Gamma^{-1})_{jj} = 1-\frac{1}{t_d} \tau_{Lj}\tau_{Rj} \end{equation}
and the reflection amplitude is
\begin{equation}
 -1+(\Gamma^{-1})_{jj}=-\frac{1}{t_d} \tau_{Lj}\tau_{Rj} \end{equation}

Note also that the $S$-matrix has to be unitary. An interesting
question is this: Which properties should the $\Gamma$ matrix
satisfy so that the resultant $S$-matrix is unitary? It appears
that the following equations
\begin{eqnarray}
  \varphi_L(x)^* = r_0^* \varphi_L(x)+ t_0^* \varphi_R(x)\quad,\\
  \varphi_R(x)^* = t_0^* \varphi_L(x)+ r^{\prime*}_0 \varphi_R(x)\quad,
\end{eqnarray}
which are also satisfied by $f_L$ and $f_R$, are the only
ones we need. From here, it can be shown that $\Gamma$-matrix
and its inverse satisfy
\begin{eqnarray}
 \Gamma+\Gamma^\dagger -2I &=& f_L f_L^\dagger+f_R f_R^\dagger\quad,\\
 \Gamma^{-1}+(\Gamma^\dagger)^{-1} -2\Gamma^{-1}(\Gamma^\dagger)^{-1}
    &=& \tau_L \tau_L^\dagger+\tau_R \tau_R^\dagger\quad.
\end{eqnarray}
Unitarity of $S$-matrix follows from these.

\subsection{Probabilities}

We are using mostly the transmission probabilities. The direct
transmission probability is $T_d=\abs{t_d}^2$. The transmission
probability from left lead to a direction in probe-$j$ and the
corresponding quantity for the right lead are
\begin{eqnarray}
  T_{Lj} &=& \abs{\tau_{Lj}}^2\quad,\\
  T_{Rj} &=& \abs{\tau_{Rj}}^2\quad.
\end{eqnarray}
The transmission probabilities between two different probes $j$
and $\ell$ can be expressed in terms of the quantities above
\begin{equation}
 T_{j\ell} = \abs{\frac{1}{t_d}~\tau_{Rj_<}~\tau_{Lj_>}}^2
=\frac{T_{Rj_<}~T_{Lj_>}}{T_d}\quad.
\end{equation}
In other words, knowing the transmission probabilities for the
main conductor, we can determine these probabilities between the
probes.

\subsection{Conductance}

We suppose that the leads of the main conductor have electrostatic
potentials $W_L$ and $W_R$. We assume that both directions on the
probes are at the same potential $W_j$. The differences in chemical
potentials are related to these potentials by
$\mu_L-\mu_R=(-e)(W_L-W_R)$ etc.

The current that enters from the lead $\alpha$ and go to the lead
$\beta$ can be expressed as
\begin{equation}
I_{\alpha\rightarrow\beta}=2\frac{(-e)}{h}(\mu_\alpha-\mu_\beta)=\frac{2e^2}{h}(W_\alpha-W_\beta)=
G_0(W_\alpha-W_\beta)\quad,
\end{equation}
where $G_0$ is the conductance quantum. Form these we can get
expressions for the total current going into a lead.
\begin{eqnarray}
  I_L&=&G_0\left[T_d(W_L-W_R)+\sum_j  2T_{Lj}(W_L-W_j)\right]\quad,\\
  I_R&=&G_0\left[T_d(W_R-W_L)+\sum_j  2T_{Rj}(W_R-W_j)\right]\quad,\\
  I_j=I_{j^\prime} &=& G_0\left[T_{Lj}(W_j-W_L)+T_{Rj}(W_j-W_R)+
  T_{j j^\prime}(W_j-W_j)\right.\\
  & &\qquad \left.+\sum_{\ell\neq j}  2T_{\ell j}(W_j-W_\ell)\right]\quad,
\end{eqnarray}
The total current going in has to be zero: $I_L+I_R+\sum_j
2I_j=0$. Also, all the potentials $W_\alpha$ can be shifted by a
constant amount, $W_\alpha\rightarrow W_\alpha+\delta W$, and this
does not change the value of currents. Due to this we can choose
one of the potentials (such as $W_R$) to be 0 (grounding).

Since probes are only imaginary, we require them to carry no
current, $I_j=0$. In this way, if electrons go into one of these
probes, same number of electrons come back. In this way, particles
do not disappear on the average on the main conductor. In that case
we have $I_L=-I_R=I$, the current passing from the device. We
suppose that $W_R=0$ and express all other potentials in terms of
$W_L$.

The equation for the current entering into probe-$j$ is
\begin{equation}
 T_{Lj} W_L = \left(T_{Lj}+T_{Rj}+2\sum_{\ell\neq j} T_{\ell j}\right)W_j
- 2\sum_{\ell\neq j} T_{j \ell} W_\ell\quad.
\end{equation}
The terms inside the parentheses is equal to (by the unitarity of $S$-matrix)
\begin{equation}
1-\abs{S_{jj}}^2-\abs{S_{jj^\prime}}^2
=\left(\frac{\tau_{Lj}\tau_{Rj}}{t_d}\right)+\left(\frac{\tau_{Lj}\tau_{Rj}}{t_d}\right)^*
-2\frac{T_{Lj} T_{Rj}}{T_d}=m_j-2\frac{T_{Lj} T_{Rj}}{T_d}\quad.
\end{equation}
We define a new $N\times N$ matrix, $P$, with
\begin{equation}
 P_{j\ell}=m_j\delta_{j\ell}-2\frac{T_{Rj_<} T_{Lj_>}}{T_d}\quad.\end{equation}
It is a symmetric matrix with real elements which also satisfies (because
of the way the diagonal elements are defined)
\begin{equation}
 \sum_\ell P_{j\ell}=T_{Lj}+T_{Rj}\quad.\end{equation}
Using this matrix, we can find the potentials $W_j$,
\begin{equation}
 \frac{W_j}{W_L}=  \sum_\ell P^{-1}_{j\ell} T_{L\ell}\quad.\end{equation}
Using these, the dimensionless conductance can be expressed as
\begin{eqnarray}
  g &=& \frac{I}{G_0 W_L} \\
    &=& T_d+2\sum_j T_{Lj} -2\sum_{j\ell} T_{Lj} P^{-1}_{j\ell}T_{L\ell}\\
    &=& T_d + 2\sum_{j\ell} T_{Rj} P^{-1}_{j\ell}T_{L\ell}
\end{eqnarray}

\section{The continuum version}
We are now going to pass from the discrete model solved above to a
continuum model where the probes are infinite in number and they are
distributed uniformly to every position. Still, we may want to keep
a finite range for the positions where these probes are in contact
with the main conductor. For this reason, we suppose that the region
where decoherence occurs is on the interval between positions
$x_L^D$ and $x_R^D$.

Second, we are going to make a connection with the previous
discrete problem. So, we are going to select  $N$ points uniformly
within the decoherence interval.
\begin{equation}
x_L^D\leq x_1<x_2<\cdots<x_N\leq x_R^D\quad.\end{equation}
We are not going to specify how these points are chosen, but in
$N\rightarrow\infty$ limit, they should fill out the whole
interval. Let $\Delta x_j$ be the length of interval where the point $x_j$
corresponds to. A possible choice might be $\Delta
x_j=x_{j+1}-x_j$ and $\Delta x_N=x_R^D-x_N$ if $x_1=x_L^D$. Another possibility
is choosing $x_j$ in the middle of each subinterval of length
$\Delta x_j$. In all cases, we should have $\sum \Delta x_j
=(x_R^D-x_L^D)$.

We are going to define $d_j$, the coupling strength to probe-$j$,
by
\begin{equation}
d_j = d(x_j)\sqrt{\Delta x_j}\quad,
\end{equation}
where $d(x)$ is a real function defined on the decoherence
interval. It has dimensions of Energy$\times$Length$^{1/2}$.
Similarly, the potential of probe-$j$, $V_j$, has to be chosen as
a continuous function of position of contact, $x_j$. Let
$\hat{V}(x)$ denote this function, i.e., $V_j=\hat{V}(x_j)$. The
velocity at probe-$j$, $v_j$, will then be
\begin{equation}
v_j=v(x_j)=\sqrt{2(E-\hat{V}(x_j))/m^*}\quad.\end{equation} Then
we will define $D(x)$ function as
\begin{equation}
 D(x)=\frac{d(x)}{\hbar\sqrt{v(x)}}\quad,\end{equation}
and the coefficients $ D_j$ becomes $D(x_j)\sqrt{\Delta x_j}$. For
this reason, the function $D(x)$ has the dimensions of
Time$^{-1/2}$. Hopefully, we are going to demonstrate with
numerical solutions that $D(x)^2$ corresponds to the decoherence
rate $1/\tau_\phi$.

It is natural to define the two functions $f_L(x)$ and $f_R(x)$ as
\begin{equation}
f_L(x)=D(x)\varphi_L(x)\quad,\qquad f_R(x)=D(x)\varphi_R(x)\quad.
\end{equation}
In that case we have $f_{Lj}=f_L(x_j)\sqrt{\Delta x_j}$ and
$f_{Rj}=f_R(x_j)\sqrt{\Delta x_j}$. (The functions $f_{L,R}(x)$
have the dimensions Length$^{-1/2}$, but $f_{L,Rj}$ are
dimensionless.)

The $\Gamma$ matrix is defined in the usual way as
\begin{equation}
 \Gamma_{j\ell}=\delta_{j\ell}+\frac{1}{t_0}f_{Rj_<}f_{Lj_>}
=\delta_{j\ell}+\frac{1}{t_0}f_R(x_{j_<})f_L(x_{j_>})\sqrt{\Delta x_j\Delta x_\ell}
\end{equation}
We are interested in obtaining a functional form for the $\Gamma$ matrix. Note that in discrete form,
$\Gamma^{-1}$ is applied to the vectors  which have $\sqrt{\Delta x}$ factors in all of
their elements. For this reason, let us investigate the general
relation $a_j=\Gamma_{j\ell}~b_\ell$ where $a_j=a(x_j)\sqrt{\Delta x_j}$ and
$b_j=b(x_j)\sqrt{\Delta x_j}$.
\begin{equation}
 a(x_j)\sqrt{\Delta x_j}= b(x_j)\sqrt{\Delta x_j}+\frac{\sqrt{\Delta x_j}}{t_0}
\sum_\ell f_R(x_<) f(x_>) ~ b(x_\ell)\Delta x_\ell\quad.
\end{equation}
Eliminating the common factors in square roots we get a functional equation
\begin{equation}
 a(x)=\int \Gamma(x;y) b(y) dy \quad,\end{equation}
where
\begin{equation}
  \Gamma(x;y)=\delta(x-y) +\frac{1}{t_0} f_R(x_<) f_L(x_>)\quad.
\end{equation}

Therefore we are going to define functions $\tau_L(x)$ and
$\tau_R(x)$ (defined only on the decoherence interval) by
\begin{equation}
 f_{L,R}(x)=\int \Gamma(x;y) \tau_{L,R}(y) dy\quad.
\end{equation}
Using these we have $\tau_{Lj}=\tau_L(x_j)\sqrt{\Delta x_j}$ etc.
Similarly the inverse of $\Gamma$ function can be expressed as
\begin{equation}
\Gamma^{-1}(x;y)=\delta(x-y) -\frac{1}{t_d} \tau_R(x_<)
\tau_L(x_>)\quad,
\end{equation}
where
\begin{equation}
t_d=t_0 -\sum_j \tau_{Rj} f_{Lj}=t_0-\int \tau_R(x) f_L(x)
dx\quad.\end{equation} The reflection amplitudes can also be
expressed in the same form.

The transmission probabilities are
\begin{equation}
T_{Lj}=\abs{\tau_L(x_j)}^2 \Delta x_j =T_L(x_j) \Delta
x_j\quad,\qquad T_{Rj}=\abs{\tau_R(x_j)}^2 \Delta x_j =T_R(x_j)
\Delta x_j\quad.
\end{equation}
It is good that the probabilities turn out to be proportional to
the interval length (Note that the probe-$j$ takes care of the
decoherence on an interval with length $\Delta x_j$). The
transmission between two different intervals
\begin{equation}
T_{j\ell}=\frac{\abs{\tau_R(x_<)}^2
\abs{\tau_R(x_>)}^2}{T_d}\Delta x_j \Delta x_\ell
=\frac{T_R(x_<)~T_L(x_>)}{T_d}\Delta x_j \Delta x_\ell
\end{equation}
is also proportional to both of the lengths of the corresponding intervals.

Next, note that
\begin{equation}
 m_j =2\real\frac{\tau_R(x_j)\tau_L(x_j)}{t_d}~~\Delta x_j = M(x_j)\Delta x_j\quad.
\end{equation}
The matrix elements of $P$ becomes
\begin{equation}
 P_{j\ell}=\delta_{j\ell} M(x_j)\Delta x_j
-2 \frac{T_R(x_<)T_L(x_>)}{T_d}\Delta x_j \Delta x_\ell\quad.
\end{equation}
This matrix looks different from $\Gamma$ in the way it contains interval
lengths. But still we can define a function form
\begin{equation}
 P(x;y)=M(x)\delta(x-y)-2 \frac{T_R(x_<)T_L(x_>)}{T_d}\quad.\end{equation}
So, if $W(x_j)$ denotes the electrostatic potential on probe-$j$,
we have
\begin{equation}
 T_L(x)=\int P(x;y) \frac{W(y)}{W_L}dy\quad.\end{equation}
$P(x;y)$ also satisfies the equation
\begin{equation}
 \int P(x;y)dy = T_L(x)+T_R(x)\quad.\end{equation}

Finally, it can be shown that the dimensionless conductance $g$ can be expressed as
\begin{eqnarray}
  g &=& \frac{I}{G_0 W_L} \\
    &=& T_d+2\int T_L(x)dx -2\int\int T_L(x) P^{-1}(x;y) T_L(y) dxdy\\
    &=& T_d + 2\int\int T_R(x) P^{-1}(x;y) T_L(y) dxdy
\end{eqnarray}
where $P^{-1}(x;y)$ is the inverse of $P(x;y)$
\begin{equation}
 \int P^{-1}(x;y)P(y;z) dy =\delta(x-z)\quad.\end{equation}

\section{Small decoherence rate}
In here we will assume that the coupling strength expression
$d(x)$ is small, so that we can expand all relevant quantities in
$D(x)$. Mostly we will be interested in the lowest order term. The
functions $f_L$ and $f_R$ are of first order in $g$. The $\Gamma$
function-matrix is
\begin{equation}
 \Gamma(x;y)=\delta(x-y)+\frac{1}{t_0} f_R(x_<) f_L(x_>)\quad,\qquad
\Gamma^{-1}(x;y)\approx \delta(x-y)-\frac{1}{t_0} f_R(x_<)
f_L(x_>)\quad.\end{equation} From here we get $\tau_L\approx f_L$
and $\tau_R\approx f_R$ where the corrections are of third order.

The direct transmission amplitude is
\begin{equation}
 t_d\approx t_0-\int f_L(x) f_R(x)dx \quad.\end{equation}
The direct transmission probability becomes
\begin{equation}
 T_d\approx\abs{t_0}^2\left(1-\int \frac{f_L(x)f_R(x)}{t_0}dx - \int \frac{f_L^*(x)f_R^*(x)}{t_0^*}dx \right)\quad.\end{equation}
Note that
\begin{equation}
 M(x)=2\real \frac{\tau_R(x)\tau_L(x)}{t_d} \approx 2\real \frac{f_R(x)f_L(x)}{t_0}\quad,\end{equation}
which is of second order, as a result we can express $T_d$ as
\begin{equation}
 T_d\approx\abs{t_0}^2\left(1-\int M(x) dx\right)\quad.\end{equation}

The transmission probability densities to the probes are
\begin{equation}
 T_L(x) \approx \abs{f_L(x)}^2\quad,\qquad T_R(x) \approx \abs{f_R(x)}^2\quad, \end{equation}
which are of second order. Therefore, the $P$ matrix-function
\begin{equation}
 P(x;y)=M(x)\delta(x-y) - \frac{2}{T_d}T_R(x_<) T_L(x_>)\quad,\end{equation}
has at least a second order term as the first term and a fourth order term in the last term.
For this reason, we might need to calculate $M(x)$ to fourth order as well. Let us consider the
problem in the following way. Write the matrix as $P=P_1+P_2$ where $P_1=M(x)\delta(x-y)$ and
$P_2$ is the remaining term. Inverse of $P$ is
\begin{equation}
P^{-1}=P_1^{-1} - P_1^{-1} P_2 P_1^{-1} + P_1^{-1} P_2 P_1^{-1} P_2 P_1^{-1}-\cdots \end{equation}
Since $P_1^{-1}=M(x)^{-1}\delta(x-y)$, we have
\begin{equation}
 \int\int T_R(x) P^{-1}(x;y) T_L(y) dx dy=\int \frac{T_R(x)T_L(x)}{M(x)}dx
  -\int \int \frac{T_R(x)P_2(x;y)T_L(y)}{M(x)M(y)}dx dy +\cdots\quad,
\end{equation}
where the first term is of second order and the second one is of fourth order. We keep the first
term only. For this reason, we don't need to calculate the higher order terms in $M(x)$. The result
for the dimensionless conductance is
\begin{equation}
 g=\abs{t_0}^2\left(1-\int M(x)dx\right)+2\int  \frac{T_R(x)T_L(x)}{M(x)}dx \quad.\end{equation}

\subsubsection*{Summary of the steps of a numerical computation}

\begin{itemize}
\item A potential $V(x)$ has to be chosen and the solutions
$\varphi_{L,R}$ of the Schr\"odinger equation at a selected energy
$E$ have to be obtained. We will use
$\tilde{\varphi}_{L,R}=\sqrt{v_L}\varphi_{L,R}$ which are
dimensionless. Through the solutions, we also obtain the scattering
matrix of the ``bare'' main conductor, the amplitudes $r_0$,
$r_0^\prime$ and $t_0$; but we need only the transmission amplitude
$t_0$. \item A decoherence interval (from $x_L^D$ to $x_R^D$) has to
be chosen and a function $\tilde{D}(x)$ has to be defined on this
interval. $\tilde{D}(x)$ has the dimensions of Length$^{-1/2}$. It
is related to $d(x)$ through the relation
$\tilde{D}(x)=d(x)/\hbar\sqrt{v(x)v_L}$. We ignore the energy
dependence of $\tilde{D}$ and for all energies, $E$, use the same
function. \item For the calculation, we divide the interval
$[x_L^D,x_R^D]$ into $N$ subintervals each with length $\Delta x_j$
and positioned at $x_j$. We choose $N$ to be large enough so that
each subinterval is smaller than the wavelength of solutions (or
smallest length scales associated with the wavefunctions
$\tilde{\varphi}_{L,R}$). \item We define $N\times1$ column matrices
$f_{Lj}$ and $f_{Rj}$ by
\begin{equation}
 f_{Lj}=\tilde{D}(x_j)\tilde{\varphi}_L(x_j)\sqrt{\Delta x_j}\quad,
\qquad f_{Rj}=\tilde{D}(x_j)\tilde{\varphi}_R(x_j)\sqrt{\Delta
x_j}\quad,
\end{equation}
\item
We construct the $\Gamma$ matrix by
\begin{equation}
 \Gamma_{j\ell}=\delta_{j\ell}+\frac{1}{t_0} f_{Rj_<} f_{Lj_>}\quad.\end{equation}
\item We obtain $N\times1$ column matrices $\tau_{Lj}$ and
$\tau_{Rj}$ by $\tau_L=\Gamma^{-1} f_L$ and $\tau_R=\Gamma^{-1}
f_R$. \item The direct transmission amplitude is calculated by using
$ t_d = t_0 -\tau_R^Tf_L$ and the associated probability by
$T_d=\abs{t_d}^2$.

\item The transmission probabilities from the left and right leads
to the probes are obtained by $T_{Lj}=\abs{\tau_{Lj}}^2$ and
$T_{Rj}=\abs{\tau_{Rj}}^2$. Also, we find $m_j$ by
\begin{equation}
 m_j=2\real \frac{\tau_{Rj} \tau_{Lj}}{t_d}\quad.\end{equation}
\item
We will define a matrix $P$ by
\begin{equation}
P_{j\ell}=m_j\delta_{j\ell}-\frac{2}{T_d} T_{Rj_<}
T_{Lj_>}\quad.\end{equation} \item The dimensionless conductance and
the local electrostatic potentials of probes are calculated by
\begin{eqnarray}
 g &=& T_d +2 T_R^T P^{-1} T_L\quad,\\
\frac{W_j}{W_L} &=& (P^{-1}T_L)_j\quad.
\end{eqnarray}

\end{itemize}

\section{Results and Conclusion}

In this work we have revealed our continuum model for decoherence in
1D transport through a mesoscopic wire. The dephasing effects in 1D
transport had been investigated by extending B\"uttiker dephasing
model, which is a conceptually simple model to simulate the
dephasing effect in 1D transport through a mesoscopic system by
coupling electron reservoir to the conductor. In our model
decoherence proceeds at every location such that we coupled 2N
electron reservoirs to the conductor by 2N channels and we choose N
to be large to obtain a continuum case. In the reservoirs inelastic
events and phase randomization take place. Electrons can go to
equilibrium in those channels but will eventually return back into
the system and at the end, as a result of dephasing, coherence is
lost, same as in the B\"uttiker's dephasing model. Our model is more
consistent with the prevalent notions of decoherence since the
placement of the single scatterer in B\"uttiker's model effects the
electron transmission.

The key point that we have solved in this work is whether extending
B\"uttiker's fictitious probe model can be made and give us more
reliable data. We apply our model of continuum decoherence for the
double barrier case in a one dimensional wire at mesoscopic scales
and focus on resonant tunneling seen in such devices.

Incident electrons are described by plane waves. We consider
potentials with $V(x\rightarrow-\infty)=V(x\rightarrow+\infty)$ so
that $k_{L}=k_{R}$ and $\upsilon_{L}=\upsilon_{R}$. In this case
$\frac{1}{\sqrt{\upsilon_{L}}}$ for $\varphi_{L}$ and $\varphi_{R}$
can be absorbed into $\gamma$, i.e.,
\begin{eqnarray*}
\tilde{\varphi}_L=\sqrt{\upsilon_{L}} \varphi_{L} &=& \left\{
\begin{array}{ll}
\left( e^{ikx} + r_{0}e^{-ikx}\right)  & \textrm{for}~~x\rightarrow-\infty\\
\left( t_{0} e^{ikx}\right) & \textrm{for}~~x\rightarrow+\infty
\end{array}
\right. \\
\tilde{\varphi}_R=\sqrt{\upsilon_{L}} \varphi_{R} &=& \left\{
\begin{array}{ll}
\left( t_{0} e^{-ikx}\right)  & \textrm{for}~~x\rightarrow-\infty\\
\left( e^{-ikx} + r_0^\prime e^{ikx}\right) &
\textrm{for}~~x\rightarrow+\infty
\end{array}
\right.
\end{eqnarray*}
$\tilde{\gamma}_{j}=\frac{\gamma_{j}}{\sqrt{\upsilon_{L}}}$ so
$f_{R,j}=\tilde{\gamma}_{j}\tilde{\varphi}_{R}(x_{j}).$ In this case
$\tilde{\varphi}_{L,R}$ and $\tilde{\gamma}$ are dimensionless.

\begin{figure}
\begin{center}
\includegraphics[scale=0.60]{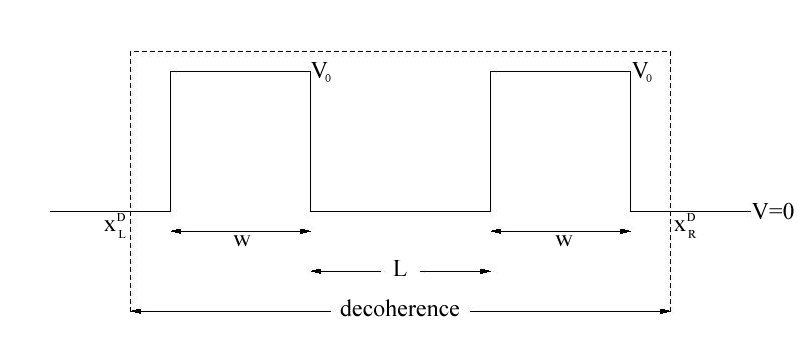}
\end{center}
\caption{Double barrier case.} \label{fig:doublebar}
\end{figure}

Electron waves tunnel through the left and right barriers via the
quantum well. The potential felt by the electrons is depicted in
Fig.~\ref{fig:doublebar}. In the well the electron wave experiences
multiple reflections due to the barriers and then the wave tunnels
out the right barrier. Transfer matrix method is used to calculate
the reflection and transmission coefficients. The barriers transfer
matrices are obtained by matching the wave functions and their
derivatives at the boundaries. So we had the transmission and
reflection amplitudes. Once we get the transmission probability we
apply our procedure to get the conductance g.

\begin{figure}
\begin{center}
\includegraphics[scale=0.80]{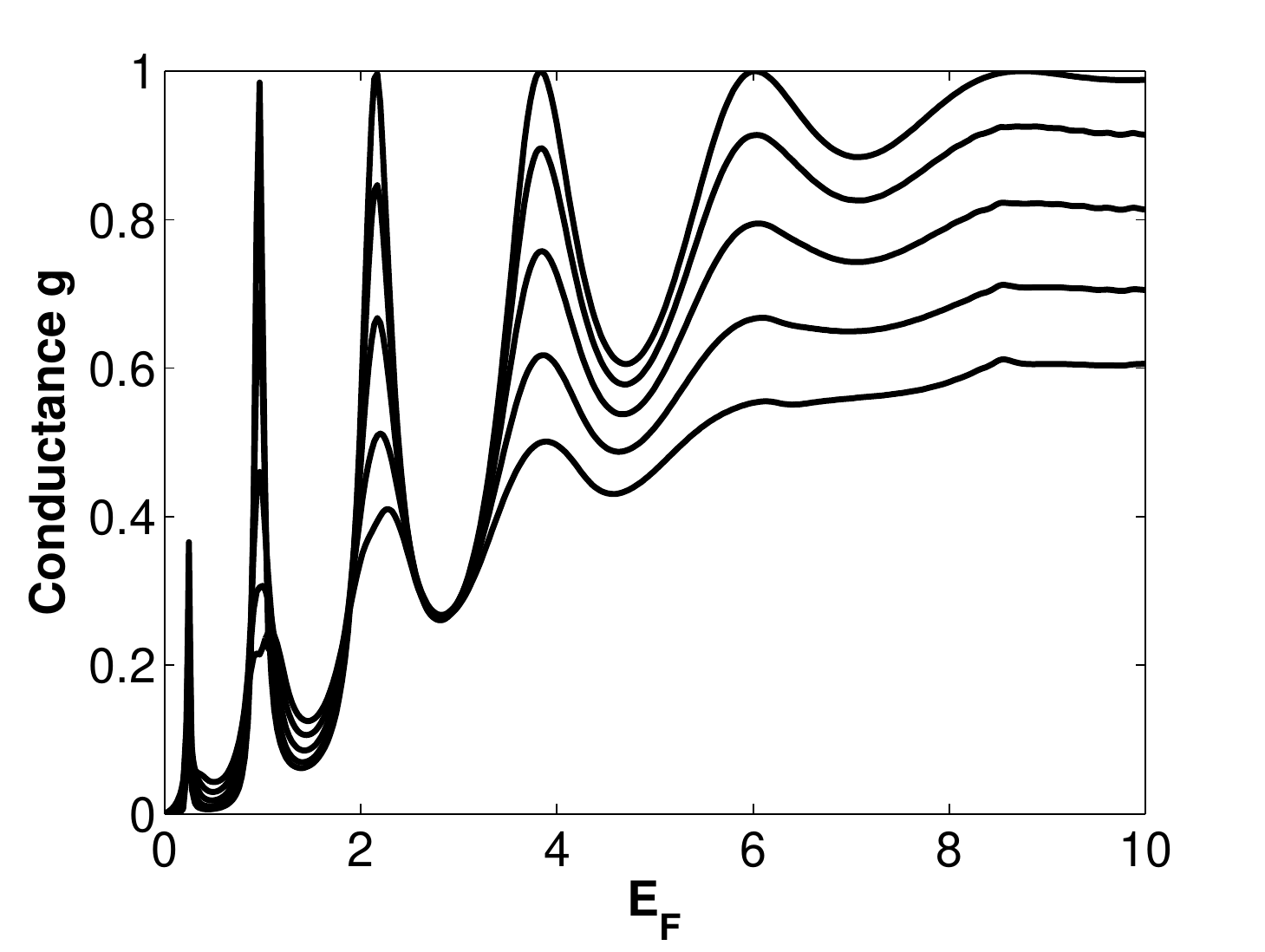}
\end{center}
\caption{Conductance vs $E_{F}$ graph for different D values. D
values are 0, 0.3, 0.5, 0.7, 0.9 from top right to bottom right.}
\label{fig:doublegvsE}
\end{figure}
\begin{figure}
\begin{center}
\includegraphics[scale=0.80]{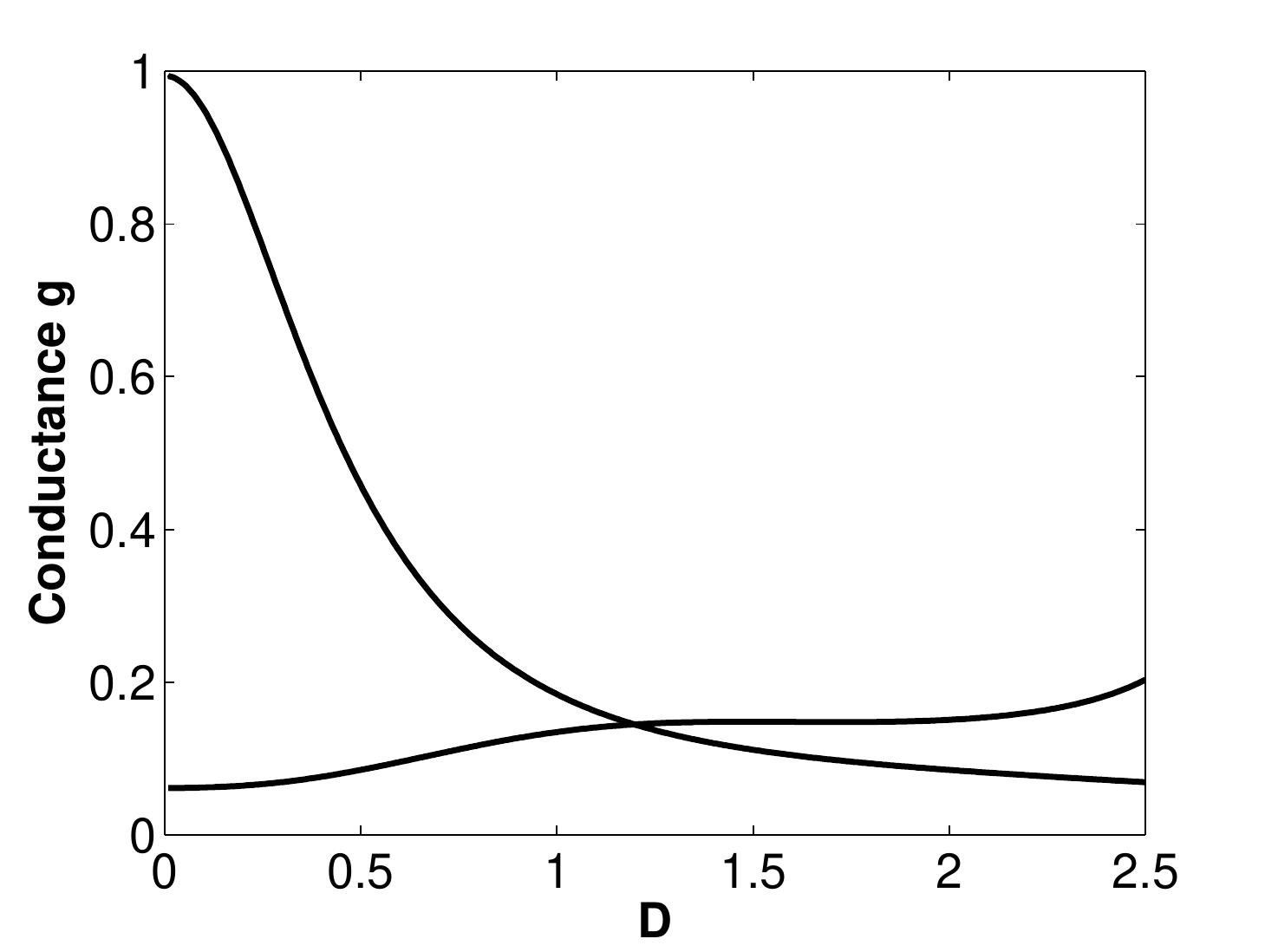}
\end{center}
\caption{Conductance vs D graph for $E_{F}=E_{1}=0.96$ which is the
second maximum at Conductance vs $E_{F}$ graph for different D
values(Fig.~\ref{fig:doublegvsE}) and for $E_{F}=E_{2}=1.41$ which
is the second minimum in the same Fig.~\ref{fig:doublegvsE}.}
\label{fig:doublegvsd}
\end{figure}

Fig.~\ref{fig:doublegvsE} shows conductance versus $E_{F}$ graph for
different D values for the double barrier case. As seen in the
figure the conductance decreases with the increase in decoherence.
D=0 case is shown at the top. The peaks seen in the tunnelling
region, where the energies are smaller than $V_{0}=2.5$, are due to
resonant transmission. In this region we see that decoherence makes
the constructive interference of electron waves disappears. After
that region we see that conductance,i.e. the electron transmission,
is suppressed by dephasing.

Fig.~\ref{fig:doublegvsd} shows conductance versus D graph for
$E_{F}=E_{1}=0.96$ and for $E_{F}=E_{2}=1.41$ which is the second
maximum and second minimum at Conductance vs $E_{F}$ graph for
different D values(Fig.~\ref{fig:doublegvsE}).

Decoherence mainly prevents wave interference. Depending on whether
the interference increase or decrease the transmission probability,
decoherence may decrease or increase the conductance. So, if
constructive interference is present in the forward direction
decoherence will prevent that and decrease the conductance.
Otherwise, if destructive interference is effective in the forward
direction, then decoherence increases the conductance. But as a
rough guide we can give the following rule: When the transmission
probability is roughly below 0.1, decoherence increases the
conductance. Otherwise, if the transmission probability is above
0.1, then decoherence decreases the conductance.

In summary, we have proposed a model to address the significant
dephasing effects in 1D transport.And we observe that dephasing can
dramatically suppress the conductance of a conductor since it
effects the transmission probability of the electron waves.


\end{document}